\documentclass[aps,prb,onecolumn,superscriptaddress]{revtex4}
\usepackage{graphicx}

\begin{document}
\title{Impurity center in a semiconductor quantum ring in the presence
of a radial electric field}

\author{Boris~S.~Monozon}
\email{monozon@mail.gmtu.ru}
\affiliation{Physics Department, Marine Technical University, 3 Lotsmanskaya Str.,
190008 St.Petersburg, Russia}

\author{Mikhail V. Ivanov}
\email{mivanov@mi1596.spb.edu}
\altaffiliation{permanent address: Institute of Precambrian Geology
and Geochronology, Russian Academy of Sciences, Nab. Makarova 2,
St. Petersburg 199034, Russia }
\affiliation{Theoretische Chemie, Institut f\"ur Physikalische Chemie,
Universit\"at Heidelberg,
INF 229, 69120 Heidelberg, Germany}

\author{Peter Schmelcher}
\email{Peter.Schmelcher@pci.uni-heidelberg.de}
\affiliation{%
Theoretische Chemie, Institut f\"ur Physikalische Chemie,
Universit\"at Heidelberg,
INF 229, 69120 Heidelberg, Germany}%
\affiliation{%
Physikalisches Institut, Universit\"at Heidelberg, Philosophenweg 12, 69120 Heidelberg, Germany}%


\date{\today}

\begin{abstract}
The problem of an impurity electron in a quantum ring (QR) in
the presence of a radially directed strong external electric field is investigated in detail.
Both an analytical and a numerical approach to the problem are developed.
The analytical investigation focuses on the regime of a strong wire-electric field compared to the
electric field due to the impurity. An adiabatic and quasiclassical approximation is employed.
The explicit dependencies of the binding energy of the impurity electron on the
electric field strength, parameters of the QR and position of the
impurity within the QR are obtained. Numerical calculations of the
binding energy based on a finite-difference method in two and three dimensions
are performed for arbitrary strengths of the electric field.
It is shown that the binding energy of the impurity electron exhibits a maximum as a function
of the radial position of the impurity that can be shifted arbitrarily by applying a corresponding
wire-electric field. The maximal binding energy monotonically increases with increasing electric
field strength. The inversion effect of the electric field is found to occur.
An increase of the longitudinal displacement of the impurity typically leads
to a decrease of the binding energy. Results for both low- and high-quantum rings are derived and discussed.
Suggestions for an experimentally accessible set-up associated with the
GaAs/GaAlAs QR are provided.
\end{abstract}

\maketitle

\section{Introduction}

During the last decade electronic and optical properties of low-dimensional
semiconductor structures have been studied extensively both experimentally and
theoretically. Along with long-known systems like quantum wells, quantum wires,
quantum dots, superlattices the novel confined structures called quantum rings
(QR) attract much attention. The QR can be viewed as a cylindrical quantum dot
consisting of an axially symmetric cavity. The unique topology of the QR leads to 
remarkable quantum phenomena. In the presence of an axially directed magnetic field
the persistent current and the oscillations of the electron energy as a function
of the magnetic flux (Aharonov-Bohm effect) were found to occur \cite{AB59}.

To common knowledge impurities and/or excitons modify considerably the
electronic, optical and kinetic properties of low-dimensional structures
such as QRs. Also these properties are strongly affected by external
magnetic and electric fields. Today, an extensive literature 
is available which traces the effects of a magnetic field on free carriers,
excitons and impurity states in the QR (see for example Lin and Guo
\cite{LG04} and Monozon and Schmelcher \cite{MS03} and references therein).
At the same time the influence of an electric field on the electronic
properties of a QR has attracted much less attention. The energy levels of free electrons
and the oscillator strengths of the  interband optical transitions  as a function
of the radii of the QR and strength of the in-plane electric field were
investigated in ref.\onlinecite{Ll09}. Barticevic \emph {et al.} \cite{Ba07} studied
theoretically the effect of the in-plane electric field on the Aharonov-Bohm
oscillations and the optical absorption in the QR in the absence of 
impurities and excitons. Effects of the eccentricity and an in-plane  electric
field on the electronic and optical properties of elliptical QRs have been
considered in Ref. \onlinecite{LW02}. Recently  the influence of the in-plane
electric field on the persistent  current in the QR coupled to a quantum wire
was studied \cite{Or21}. In contrast to refs. \onlinecite{Ll09,Ba07,LW02,Or21} the
effect of the impurity centre on the electronic states in the QR subject to an
axially directed  magnetic and radially directed  electric fields was taken
into account in Ref. \onlinecite{MS03}. The influence of the radial electric  field
on the electron was assumed to be much weaker than that of the impurity centre,
magnetic field and confinement. However the influence of a strong electric
field on the impurity states in the QR is certainly of interest. The reason for this
is that a strong  electric field induces a considerable polarization of the
spatial distribution of the carriers \cite{Ll09,LW02}. Note that this may
be used in order to modulate effectively the intensity of photocurrents and
emission of light from optoelectronic devices based  on  QR structures. The
analogous effect relating to quantum wells was reported by Mendez
\emph {et al.} \cite{Me82}. Since the problem of the impurity electron in the QR
in the presence of a strong electric field is not addressed in detail in the
literature we investigate this problem here.
In case the electric field is parallel to the symmetry axis of the QR the ring topology
is preserved but there is no significant effect on the radial states of the QR.
For a strong in-plane electric field the ring topology is broken leading to
the disappearance of the unique ring properties. It is therefore most advantageous 
to apply a radially directed electric field created by a wire whose position coincides
with the symmetry axis of the QR. In this case the electric field is directed radially
and a strong influence on the radial motion is foreseen and, equally important,
the topology of the QR is preserved.

In the present investigation of an impurity center in a semiconductor quantum ring
in the presence of radially directed electric field a two-fold approach is pursued.
First we will perform studies to obtain analytical
(approximate) solutions of the stationary Schr\"odinger equation in certain parameter
regimes. This elucidates in particular the behaviour and properties of the impurity
binding for these regimes. Second, we perform a complementary numerical investigation
that covers all possible cases (position of impurity, strength of the electric
field, radii and height of the QR). Finally, exemplary, a comparison of
analytical and computational results is provided.

In detail we proceed as follows. Section II contains a general description of the set-up
i.e. the quantum ring, the impurity and the electric field configuration.
Section III is devoted to the analytical investigation providing the method, the results
and their discussion. Section IV begins with an outline of our computational method
followed by a comprehensive discussion of our numerical results. Section V provides a
(naturally limited) comparison of the analytical and numerical results. Section VI contains
the conclusions.

\section{The Quantum Ring, Impurity and Field Configuration}\label{S:gen}

We consider a QR formed by the revolution of a rectangle around the $z$-axis.
The plane of the rectangle contains the $z$-axis. The QR is bounded
by potential barriers of infinite height at the planes $z=\pm d/2$ and
impenetrable cylindrical surfaces at internal $\rho=a$ and external
$\rho=b$ radii . The chosen model corresponds to a hard-wall confinement
potential. Alternatively, Chakraborty and Pietil\"{a}inen~\cite{CP94} proposed a
parabolic ring confinement potential determined by the radius of the ring
$\bar{\rho}$ and by the effective frequency $\Omega$. This potential has been
very effectively used in studies of QRs~\cite{CU02,BP00,vC01,GK02}. A comparison of
the above mentioned potential models is provided in ref.~\onlinecite{BP00}.

The position of the impurity center $\mathbf{r}_0$ is determined by the cylindrical coordinates
${a\leq\rho_0 \leq b\,,\; \varphi_0=0}$ and $-d/2\leq z_0\leq +d/2$.
Additionally a radially directed electric field
is provided by the field of a charged wire with linear effective
charge density $\lambda$ whose position coincides with the $z$-axis.
Furthermore we take the conduction band to be parabolic, non-degenerate and separated
from the valence band by a wide energy gap.

In the effective mass approximation the equation describing the impurity
electron possessing the effective mass $\mu$ at a position $\mathbf{r}(\rho,\varphi, z)$
subject to the axially symmetric and radially directed electric field has the
form
\begin{eqnarray}
\label{E:basicEq}
\left\{-\frac{\hbar^2}{2\mu}\left(\frac{1}{\rho}\frac{\partial}{\partial\rho}\rho
\frac{\partial}{\partial\rho}
+
\frac{1}{\rho^2}\frac{\partial^2}{\partial\varphi^2}
+
\frac{\partial^2}{\partial z^2}\right)
+
\frac{e\lambda}{2\pi\varepsilon_0\varepsilon} \ln\frac{\rho}{a}
-\frac{e^2}{4\pi\varepsilon_0\varepsilon
\left[ \rho^2 - 2\rho\rho_0\cos\varphi + \rho_0^2 + (z\!-\!z_0)^2 \right]^{1/2}}
\right\}\Psi(\rho,\varphi, z)
\nonumber\\
=
E\,\Psi(\rho,\varphi,z) \quad\,
\end{eqnarray}
where $\varepsilon$ is the dielectric constant.

By solving this equation subject to the boundary conditions
\begin{equation}
\label{E:Cond}
\Psi(\rho,\varphi,z)=0 \quad \mbox{for} \quad \rho=a\: ,
\ \rho=b \quad \mbox{and}\quad z=\pm\, d/2
\end{equation}
the total energy $E$ and the wave function $\Psi$ can be found in principle.

\section{Analytical Method and Results}

\subsection{Adiabatic Approach}

We assume that the effects of the lateral (within the $x$-$y$ plane) confinement and of the
electric field on the (bound) electron are taken to be much stronger than the influence of the
Coulomb field of the impurity center.
Under this condition the motion of the
electron parallel to the $z$-axis is adiabatically slower than the motion
in the $x$-$y$ plane and the cylindrical variables $\rho,\varphi$ and $z$ can be
adiabatically separated. In the adiabatic approximation the wave function $\Psi$
can be written in the form
\begin{equation}\label{E:thirdEq}
\Psi(\rho,\varphi,z)=\Theta_{N,m}(\rho,\varphi)f^{N,m}(z)
\end{equation}
where the function
\begin{equation}\label{E:forth}
\Theta_{N,m}(\rho,\varphi)=\frac{\exp({\rm i}m\varphi)}{\sqrt{2\pi}}\; R_{N,m}(\rho)
\end{equation}
describes the lateral motion of the electron of energy $E_{\perp N,m}$ determined by
the radial confinement and the electric field. $R_{N,m}(\rho)$ is the
$N$th radial state function $(N=1, 2,\ldots )$ possessing the
angular quantum number $m = 0, \pm 1, \pm 2, \ldots$. It vanishes at
$\rho = a$ and $\rho = b$ (see~eq.(\ref{E:Cond})). The function $f^{N,m}(z)$
describes the longitudinal motion parallel to the $z$-axis and
satisfies the equation
\begin{equation}\label{E:eqz}
-\frac{\hbar^2}{2\mu}\:\frac{{\rm d}^2}{{\rm d} z^2}\: f^{(N,m)}(z)
+
V_{N,m}(z)\,f^{(N,m)}(z)
=
W^{(N,m)} f^{(N,m)}(z)
\end{equation}
with the boundary conditions
\begin{equation}\label{E:condz}
f^{(N,m)}(\pm d/2)=0
\end{equation}
and with the adiabatic potential
\begin{equation}\label{E:adiabpot}
V_{N,m}(z)
=
-\frac{e^2}{4\pi\varepsilon_0\varepsilon}\int\frac{{\rm d}\mathbf{\rho}}{2\pi}\;
\frac{\vert R_{N,m}(\rho)\vert^2}{\left[\rho^2-2\rho\rho_0\cos\varphi+\rho_0^2
+(z-z_0)^2\right]^{1/2}}
\end{equation}

The binding energy $E_b=E^{(0)}-E$ of the impurity is defined as usual by
the difference between the energy of the free electron in the QR
$E^{(0)}=E_{\perp N,m}+\hbar^2\pi^{2}l^2/2\,\mu\,d^2$, $l=1,2,\ldots$ and the
energy of the impurity electron $E=E_{\perp N,m}+W^{(N,m)}$  which yields
\begin{equation}\label{E:binden}
E_b = \frac{\hbar^2\pi^{2}l^2}{2\,\mu\,d^2}
\,-\, W^{(N,m)} \ ; \quad l=1,2,\ldots
\end{equation}
where the energy of the longitudinal state $W^{(N,m)}$ is obtained by solving
eq.(\ref{E:eqz}).

\subsection{The Quasiclassical Lateral States of the Confined Electron in the
Presence of the Electric Field}\label{S:quasicl}

Let us first consider only the electron in the quantum ring i.e. we omit
the Coulomb term and the kinetic energy in $z$-direction in eq.(\ref{E:basicEq}).
Substituting
\begin{equation}\label{E:qfirst}
R_{N,m}(\rho) = \rho^{-1/2}\; \exp{({x}/{2})}\; u_{N,m}(x)
\end{equation}
where
\begin{eqnarray}\label{E:qnotate}
\rho=\rho_2\exp{x} \: ; \quad
\rho_2=a\:\! \exp(-x_1) \: ; \quad
x_1=-\frac{k^2}{s}\, ; \;
\nonumber\\
k^2=\frac{2\:\!\mu E_{\perp N,m}}{\hbar^2} \: ; \quad
s=\frac{2\:\!\mu\:\! e\lambda}{\hbar^{2}2\pi\varepsilon_0\varepsilon}
\qquad\qquad\qquad\quad
\end{eqnarray}
we obtain the equation for the function $u_{N,m}(x)$
\begin{equation}\label{E:qu}
u^{\prime\prime}_{N,m}(x) -
\Bigl( m^2 + s\:\! \rho_2^2\,x\,\exp{(2x)} \Bigr)\, u_{N,m}(x) = 0
\end{equation}
The boundary conditions (\ref{E:Cond}) become
$u(x_1)=u(x_2)=0$, $x_2=\ln\left(b/a\right)+x_1$. Further we consider
the cylindrically symmetric radial state only and the corresponding $m=0$ label will
be dropped in the following.

Since equation (\ref{E:qu}) does not allow for an exact analytical solution
for arbitrary magnitudes of the electric field $\propto s$ a quasiclassical approach
will be used. This method was developed originally for a free electron in
an unbound medium in the presence of a radial electric field in ref.~\onlinecite{GP78}.
In case of a positively charged wire $(s>0)$ attracting the electron to the
internal surface of the QR the wave function $u_N(x)$ is given by the equation
\begin{equation}\label{E:qufunct}
u_N(x)=\left(\frac{4\:\!s}{\pi}\right)^{\!1/4}
Q^{-1/2}(x)\: \cos\!\left[F(x)-\frac{\pi}{4}\right]
\end{equation}
where $Q(x)=\rho_2\, s^{1/2}\:\! (-x)^{1/2} \exp(x)$ and where
\begin{equation}\label{E:qFfunct}
F(x)=\int_{x_1}^x Q(t)\, {\rm d}t
\end{equation}

The parameters $k$ and $\rho_2$ are determined by the Bohr-Sommerfeld quantization
rule
\begin{equation}\label{E:qBZom}
\int_{x_1}^0 Q(t)\, {\rm d}t = \pi\:\! (N+{1}/{2}) \; ; \quad N=0,1,2,\ldots
\end{equation}

The limits of the integration in eq.(\ref{E:qBZom}) $x=x_1$ and $x=0$ correspond
to the near and far turning points $\rho=a$ and $\rho=\rho_2$. Note that
the distance $\rho_2$ determines the region of the localization of the electron
density. Under the condition $-x_1\gg1$ the integration in eq.(\ref{E:qBZom}) can be
performed explicitly thereby providing the expressions for the energy of
the lateral motion $E_{\perp N}$ \cite{GP78} and the radius of the
$N$th radial state $\rho_{2N}$
\begin{equation}\label{E:qEperp}
E_{\perp N}=\frac{\hbar^2 s}{2\:\!\mu}\,
\ln\left(\frac{\rho_{2N}}{a}\right)\ ; \quad
\quad\rho_{2N}=\left(\frac{4\pi}{s}\right)^{\!1/2} (N+{1}/{2})
\end{equation}
as well as the function $F(x)$ (\ref{E:qFfunct}) for $x_1 \leq x \leq 0$
\begin{equation}\label{E:qF1}
F(x)=\pi\:\! (N+{1}/{2})
\left( 1 -
\left[ \Phi(\sqrt{-x}) - \frac{2}{\pi^{1/2}}\,\sqrt{-x}\,\exp{x} \right]
\right)
\end{equation}
where $\Phi(t)$ is the probability integral~\cite{AS}. Eqs.(\ref{E:qEperp}),
(\ref{E:qF1}) are valid under the conditions
\begin{equation}\label{E:qineq}
\frac{a}{\rho_{2N}}\ll 1 \; ;\qquad \ln\left(\frac{\rho_{2N}}{a}\right) \gg 1
\end{equation}

For the narrow QR satisfying the conditions
$$
(b-a)\ll a,b \; ;\qquad \frac{(b-a)^3\, s}{2\pi^2\left(N+1\right)^2 a}\ll 1
$$
the Bohr-Sommerfeld quantization rule
$$
\int_{x_1}^{x_2} Q(t)\, {\rm d}t = \pi\left(N+1\right)\; ; \quad N=0,1,2,\ldots
$$
leads to the energy of the lateral motion
\begin{equation}\label{E:qEn}
E_{\perp N}=\frac{\hbar^2\pi^2\left(N+1\right)^2}{2\:\!\mu\:\!(b-a)^2}
+
\frac{e\lambda(b-a)}{4\pi\varepsilon_0\varepsilon\,a} \ ;
\end{equation}

This result coincides completely with that obtained in ref.~\onlinecite{MS03} where
the Schr\"{o}dinger equation was solved by means of perturbation theory. The
energy $E_{\perp N}$ (\ref{E:qEn}) is the size-quantized energy level in the
2D quantum well of width $(b-a)$ perturbed by the homogeneous electric field
with the effective strength ${\lambda\:\!(4\pi\varepsilon_0\varepsilon\,a)^{-1}}$.

For the negatively charged wire $(\lambda < 0)$ repulsion of the electron towards
the external cylindrical surface of the QR leads to the Bohr-Sommerfeld rule
$$
\int_0^{x_2}\vert Q(t)\vert\, {\rm d}t
=
\pi\left(N+{1}/{2}\right) \; ; \quad N=0,1,2,\ldots
$$
which yields for the lateral energy $E_{\perp N}$ the result
\begin{equation}\label{E:qEperp1}
E_{\perp N}
=
-\frac{\hbar^2}{2\:\!\mu}\: |s|\, \ln\!\left(\frac{b}{a}\right)
+
\frac{\hbar^2}{2\:\!\mu}
\left[ \frac{3\:\!\pi \left(N+{1}/{2}\right) |s|}{2\:\! b} \right]^{2/3}
\end{equation}

This result is valid under the condition
$$
\frac{2\:\!\pi^2\left(N+{1}/{2}\right)^2}{|s|\, b^2} \ll 1
$$

It follows from above that the energy $E_{\perp N}$ (\ref{E:qEperp1}) is the sum of
the lowest potential energy of the electron positioned at $\rho=b$ in the presence
of the radial electric field and a perturbatively acting QR confinement.

Next we consider the positively charged wire $(\lambda>0)$ and
relatively wide QR for which the lateral energy $E_{\perp N}$ is given by
eq.(\ref{E:qEperp}). In parallel with this we assume that the condition
\begin{equation}\label{E:qend}
\rho_{2N}\ll a_0
\end{equation}
where ${a_0=4\pi\varepsilon_0\varepsilon\,\hbar^2\mu^{-1}e^{-2}}$ is the Bohr
impurity radius, holds. This means that the effect of the electric
field on the electron considerably exceeds the influence of the impurity center, so that
the energy of the lateral motion $E_{\perp N}$ of the impurity electron is
determined by the right-hand side of eq.(\ref{E:qEperp}).

\subsection{The Binding Energy of the Impurity Electron}\label{S:bind}
\subsubsection{The 2D Impurity States}

Under the condition
\begin{equation}\label{E:bineq}
d\ll a_0
\end{equation}
the states of the impurity electron have two-dimensional (2D) character.
In this case the total energy $E$ can be written in the form
\begin{equation}\label{E:bE}
E=E_{\perp N}+\Delta E_N
\end{equation}
where $E_{\perp N}$ is given by eq.(\ref{E:qEperp}) and where $\Delta E_N$ is
determined by the matrix element of the Coulomb term on the left-hand side
of eq.(\ref{E:basicEq}) in which we neglect the dependence on the coordinate $z$.
The matrix element is calculated with respect to the functions (\ref{E:qfirst}),
(\ref{E:qufunct}), (\ref{E:qF1}) with the result
\begin{equation}\label{E:bdelta}
\Delta E_N = -\:\! 4\:\! Ry\left(\frac{a_0}{\rho_{2N}}\right)
\left[\frac{1}{\pi}\, \ln\!\left(\frac{\rho_{2N}}{a}\right) \right]^{1/2}
\left[1-\frac{\ln{({\rho_0}/{a})}}{2\ln{({\rho_{2N}}/{a})}}\right]
\end{equation}
where ${Ry=\hbar^2 / 2\:\!\mu\,a_0^2}$ is the impurity Rydberg constant.
Eq.(\ref{E:bdelta}) is valid
under the conditions (\ref{E:qineq}), (\ref{E:bineq}), (\ref{E:qend})
and $a\ll\rho_0\ll\rho_{2N}$. The energy $E$ (\ref{E:bE}) is obtained by shifting the energy
of the lateral motion of the free electron in the QR $E_{\perp N}$(\ref{E:qEperp})
by the amount $\Delta E_N$ (\ref{E:bdelta}) towards lower values of the energy.
The binding energy ${E_b=E_{\perp N}-E}$ where $E$ is given by eq.(\ref{E:bE})
becomes $E_b=-\Delta E_N$ where $\Delta E_N$ is defined by eq.(\ref{E:bdelta}).

\subsubsection{The 3D Impurity States}

For the relatively high QR satisfying the conditions
\begin{equation}\label{E:bineq1}
\rho_{2N}\ll d, a_0
\end{equation}
the motion of the electron parallel to the $z$-axis described by eq.(\ref{E:eqz})
should be taken into account. The analysis of eqs.(\ref{E:eqz})--(\ref{E:adiabpot})
is based upon the Hasegawa-Howard method developed originally in ref.~\onlinecite{HH61}
and was worked out in further detail in ref.~\onlinecite{MZ91}. The details of the
application of this method to the problem of the impurity in the QR without field
and in the presence of a strong magnetic field can be
found in ref.~\onlinecite{MS03}. It allows us to restrict ourselves to the
transcendental equation for the quantum number determining the energy of the
ground state of the longitudinal motion $W$.

\paragraph{High QR $(d>a_0)$}

The equation for the quantum number $n<1$ determining the energy $W=-Ry/n^2$
has the form
\begin{equation}\label{E:btrans}
2\:\!C + \psi\, (1-n) + \frac{1}{2n} +
\ln\!\left(\frac{\rho_{2N}}{a_0\:\! n}\right)
- \Delta(\rho_0) - {1\over 2}\:\Bigl[\:\! G_1(z_0) + G_2(z_0) \:\!\Bigr] = 0
\end{equation}
where $C$ is the Euler constant ($\simeq0.577$) and where $\psi(x)$ is the
psi-function, the logarithmic derivative of the gamma function $\Gamma(x)$.
The dependencies of the energy $W$ on the radial and longitudinal positions
of the impurity center are given by the functions
\begin{widetext}
\begin{equation}
\label{E:bdelta1}
\Delta(\rho_0)
=
\cases{
\Phi(y_0^{1/2})
\left(1/2\!-\!y_0\right) + y_0 - \left(y_0/\pi\right)^{1/2}
\exp{(-y_0)} &at\ \ $\rho_0\leq\rho_{2N}$ \cr
y_0 &at\ \ $\rho_0\geq\rho_{2N}$
}
\end{equation}
\end{widetext}
where $\displaystyle{y_0=-\ln\left(\frac{\rho_0}{\rho_{2N}}\right)}$ and
\begin{equation}\label{E:bG}
G_{1,2}=\Gamma(-n)\left(\frac{d}{a_0 n}\right)^{2n}
\left\{\exp\left[\frac{d}{a_0 n}\left(1\mp\frac{2z_0}{d}\right)\right]-
1\right\}^{-1}
\end{equation}
respectively. Eq.(\ref{E:bG}) is valid under the condition
\begin{equation}\label{E:bineq2}
\left(\frac{d}{a_0 n}\right)^{\!2n} 4 \sinh\left(\frac{2\:\!z_0}{a_0 n}\right)
\left[ \exp\!\left(\frac{d}{a_0 n}\right) -
2\cosh\left(\frac{2z_0}{a_0 n}\right) \right]^{-1}\ll1
\end{equation}

Note that eq.(\ref{E:bineq2}) does not significantly limit the displacement
of the impurity $z_0$ from the symmetric plane of the QR $(z=0)$. For the QR
of height ${d\simeq 2\:\!a_0}$ subject to the radial electric field providing the
relationship ${\rho_{2N}\simeq 0.4\:\!a_0}$ and for the impurity being positioned close to the
mid plane $(z_0=0)$ and the internal surface $(\rho_0\simeq a\: ,\ y_0\gg1)$, equation
(\ref{E:btrans}) yields for the quantum number $n\simeq0.5$. Even though the
impurity is shifted by a considerable distance $z_0=d/4$ the term on the
left-hand side of eq.(\ref{E:bineq2}) is about $0.3$. In principle eq.(\ref{E:btrans})
can be solved numerically for arbitrary values of the height of the QR
$d>a_0$, and the impurity position $\rho_0$, $ d/2 -\vert z_0\vert\gg\rho_{2N}$.
However, the explicit dependencies of the energy $W$ on the above mentioned
parameters can be found for the limiting cases of small displacements $z_0$
from the symmetric plane of the QR $(2z_0/d\ll1)$, for $a\leq\rho_0\ll\rho_{2N}$
and a maximum $(\rho_0\simeq b\gg\rho_{2N})$ shift $\rho_0$ of the impurity from
the symmetric axis $\rho=0$.

For small displacements $2z_0/a_0 n\ll1$ the dependence of the quantum
number $n(z_0)<1$ and the energy $W$ of the ground state as a function of the
displacement $z_0$ can be found explicitly from eq.(\ref{E:btrans}) with the
result
\begin{equation}\label{E:bW}
W(z_0,\rho_0)
=
-\frac{Ry}{n_1^2}
\left[ 1 - 2\,\Gamma(1-n_1)
\left(\frac{2z_0}{a_0 n_1}\right)^{\!2}\left(\frac{d}{a_0 n_1}\right)^{\!2 n_1}
\exp{\!\left(-\frac{d}{a_0 n_1}\right)}
\right]
\end{equation}
where $n_1$ is the solution to eq.(\ref{E:btrans}) for $z_0=0$ and for any
radii $\rho_0$.

The effect of the radial displacement $\rho_0$ is described by the function
$\Delta(\rho_0)$ (\ref{E:bdelta1}). For $\rho=\rho_{2N}$ we have $\Delta=0$ and
with decreasing (increasing) $\rho_0$ the function $\Delta(\rho_0)$
increases (decreases) towards the internal (external) boundary of the QR. We
obtain from eq.(\ref{E:bdelta1})
\begin{equation}\label{E:bdelta2}
\Delta(\rho_0)
=
\cases{
\displaystyle{
{1\over 2}
}
\!&for\ \, $(\rho_0-a)\!\ll\! a$ \cr
\displaystyle{
\frac{\rho_{2N}\!-\!\rho_0}{\rho_{2N}}
}
\!&for\ \, $|\rho_0\!-\!\rho_{2N}|\!\ll\!\rho_{2N}$
}
\end{equation}

For small radial displacements from $\rho_0=\rho_{2N}$ for which
${4\:\!n\:\!\Delta\ll 1}$
(see below eq.(\ref{E:bdelta2})) eq.(\ref{E:btrans}) gives the approximate
expression for the quantum number $n$ and then for the energy of the ground
state $W$
\begin{equation}\label{E:bW1}
W(z_0,\Delta)=-\frac{Ry}{n_2^2}
\left[1+\frac{4\:\!n_2}{1+2n_2}\;\Delta(\rho_0)\right]
\end{equation}
where $n_2$ is the solution to eq.(\ref{E:btrans}) for $\Delta(\rho_{2N})=0$
and any positions $z_0$.

In the logarithmic approximation ($\rho_{2N}/a_0\ll 1$,
$|\ln\left(\rho_{2N}/a_0\right)|\gg 1$) the quantum number $n$ can be
calculated from eq.~(\ref{E:btrans}) explicitly
\begin{equation}\label{E:bn}
\frac{1}{n}
\,=\,
2\left\{-\ln\left(\frac{\rho_{2N}}{a_0}\right) + \Delta(\rho_0)
+
\frac{1}{2}\, \Bigl[\:\! G_1(z_0) + G_2(z_0) \Bigr] \right\}
\end{equation}

It follows from eq.(\ref{E:binden}) that the binding energy $E_b$ of the
impurity electron in the high QR can be written in the form
\begin{equation}\label{E:bbind1}
E_b=Ry\left[\left(\frac{\pi a_0}{d}\right)^2+\frac{1}{n^2}\right]
\end{equation}
where the quantum number $n$ can be found from eq.(\ref{E:btrans}) to give
particularly in the above mentioned logarithmic approximation the
expression~(\ref{E:bn}). It enables us to investigate qualitatively the dependence of the binding energy
$E_b$ on the internal radius $a$ and height $d$ of the QR, the strength of the
electric field $s$, and the position of the impurity center $z_0, \rho_0$. We
emphasize that the logarithmic approximation is used only for a qualitative
analysis.

\paragraph{Low QR ($d<a_0$)}

For positive energies $W=Ry/s^2$ in eq.(\ref{E:eqz}) using the procedure
presented in detail in ref.~\onlinecite{MS03} we obtain the transcendental equation
for the quantum number~$s$
\begin{equation}\label{E:btrans1}
\tilde{\varphi}(s) + \ln\left(\frac{\rho_{2N}}{a_0 s}\right) -
\Delta(y_0) + 1 - \frac{1}{2}
\left[\:\! \tilde{G}_1(z_0)+\tilde{G}_2(z_0) \right] = 0
\end{equation}

In eq.(\ref{E:btrans1}) the following notations are employed
$$
\tilde{\varphi}(s)=\frac{\tilde{\Gamma}(s)}{2{\rm i}}
\left\{\frac{1}{\Gamma({\rm i} s)}\left[\frac{{\rm i} \pi}{2} + 2\:\!C - 1 +
\psi(1+{\rm i} s)-\frac{1}{2{\rm i} s}\right] -\, {\rm c.c.}\right\}
$$
$$
\frac{1}{\tilde{\Gamma}(s)}=\frac{1}{2{\rm i}}
\left[\frac{1}{\Gamma({\rm i} s)}-\frac{1}{\Gamma(-{\rm i} s)}\right]
$$
$$
\tilde{G}_{1,2}(z_0)=\tilde{\Gamma}(s)\;
\frac{{\rm Re}\: W_{{\rm i} s, 1/2}(\tau_{1,2})}
     {{\rm Im}\: M_{{\rm i} s, 1/2}(\tau_{1,2})} \; ;\quad
\tau_{1,2}=\frac{2}{{\rm i}\:\! a_0\:\! s}\left(\frac{d}{2}\mp z_0\right)
$$

Eq.(\ref{E:btrans1}) can be solved explicitly for $s\ll 1$ to give for the
energy of the ground state $W$
\begin{equation}\label{E:bW2}
W=\frac{\hbar^{2}\pi^2}{2\:\!\mu\, d^2} + \Delta W
\end{equation}
where
\begin{eqnarray}\label{E:bdelW}
\Delta W
=
-Ry \left(\frac{a_0}{d}\right)
\Biggl\{
-\:\! 4 \left[\:\! C-\Delta(\rho_0) +
\ln\left(\frac{\pi\rho_{2N}}{d}\right) \right]
\cos^2\left(\frac{\pi z_0}{d}\right)\;
+\; \ln\left[\pi^2\!\left(1-\frac{4\:\!z_0^2}{d^2}\right)\right]
\Biggr\}
\end{eqnarray}

Eqs.(\ref{E:bW2}),(\ref{E:bdelW}) are valid under the
conditions~(\ref{E:bineq1}) and $d\ll\pi a_0$.
The energy $W$ (\ref{E:bW2}) is the size-quantized
ground energy level of the electron in the quantum well of width $d$
shifted towards lower energies by an amount of $\Delta W$ (\ref{E:bdelW})
associated with the impurity field. Substituting the energy $W$ (\ref{E:bW2})
into the right-hand side of eq.(\ref{E:binden}) taken for $l=1$ we obtain for
the binding energy $E_b$ of the impurity electron in the low QR the result
$E_b=-\Delta W$, where $\Delta W$ is given by eq.(\ref{E:bdelW}).

\subsection{Discussion of the Analytical Results}\label{S:disc}

\subsubsection{The 2D Impurity States.}

The binding energy $E_b$ of the 2D states has the form $E_b =-\Delta E_N$ where
the correction to the lateral energy caused by the impurity attraction is given
by eq.(\ref{E:bdelta}). It is clear from eq.(\ref{E:bdelta}) that the binding energy
decreases with increase of the internal radius of the QR $a$. Also the binding
energy decreases with increasing the radial displacement $\rho_0$ of the
impurity center from the internal surface of the QR in the region $a\ll\rho_0\ll\rho_{2N}$. Since the strong
electric field concentrates the electron density close to the internal surface
$\rho\simeq a$ the greater the distance $\rho_0-a$ between this surface and
the impurity center is the less the impurity attraction i.e. is the less the
binding energy. For the impurity positioned in the region $a\ll\rho_0\ll\rho_{2N}$
the binding energy increases with increasing electric field strength $\propto s$.
The increasing electric field shifts the electron density distributed between
$\rho\simeq a$ and $\rho\simeq\rho_{2N}$ towards the impurity center positioned
at $\rho_0\ll\rho_{2N}$. This leads to an increase in binding energy.

\subsubsection{3D Impurity States, High QR.}

The binding energy $E_b$ of the 3D states in the high QR $(d>a_0)$ is provided by
eq.(\ref{E:bbind1}). The corresponding quantum number $n$ can be found in principle
from eq.(\ref{E:btrans}). In the logarithmic approximation the quantum number
$n$ is determined by eq.(\ref{E:bn}). Since the contribution of the second term on the
right-hand side of eq.(\ref{E:bbind1}) is exponentially small compared to the
size-quantized energy $(\sim 1/d^2)$ (see eq.(\ref{E:bW})) the binding energy
decreases with increase of the height of the QR $d>a_0$. Clearly from
eqs.(\ref{E:bbind1}) and~(\ref{E:bn}) we see that the binding energy increases with
increasing the electric field. The greater the electric field is the less
the effective radius of the lateral motion $\rho_{\rm eff}\sim\rho_{2N}$ and the
greater the depth of the one-dimensional potential $V_{N, 0}$ (\ref{E:adiabpot})
governing the longitudinal motion between the bottom and top of the QR. This
leads to an increase with respect to the binding energy $E_b$.
Expressions~(\ref{E:binden}) and~(\ref{E:bW}) show that the impurity being positioned at
the mid-plane $z=0$ produces the greatest binding energy. The shift of the
impurity center $z_0$ from the mid-plane $z=0$ causes a decrease of the binding
energy $E_b$. Narrowing the QR or increasing the electric field strength increases the
energy shift $\Delta E_b(z_0)$ associated with the displacement
of the impurity from the plane $z_0=0$ in both cases.

The radial shift of the impurity center $\rho_0$ from the some intermediate
position $\rho_{0m}$
towards the internal $\rho=a$ and external $\rho=b$ surfaces  produces similar effects as those
induced by the displacement $z_0$ from the mid-plane $z=0$. It follows from eqs.
(\ref{E:bbind1}),(\ref{E:bn}),(\ref{E:bdelta1}),(\ref{E:bdelta2}) that for 
QRs possessing an external radius $b$ comparable to the impurity Bohr radius $a_0$
the binding energy $E_b (\rho_0)$ at $\rho_0=a$ approximates that of $\rho_0=b$
both being determined by the quantum number
$n\simeq\left(-2\ln\frac{\rho_{2N}}{a_0}\right)^{-1}$. Since the parameter
$\Delta (\rho_0)$ (\ref{E:bdelta2}) and quantum number $n^{-1}$ (\ref{E:bn})
both increase with the displacement of the position $\rho_0$ from the
external surface $\rho_0=b$ towards the internal one $\rho_0=a$ we expect that
the binding energy $E_b (\rho_0)$ reaches a maximum at a certain radial
position $\rho_{0m}$. The shift of the binding energy induced
by a radial displacement $\rho_0$ possesses a maximum for the impurity positioned at the plane $z_0=0$ and
decreases with increasing displacement from this plane.

It follows from the above that the corrections to the binding energy induced
by the displacements from, say, the circle $\rho_0\simeq\rho_{2N}$, $z_0=0$ to
the region $a\leq\rho_0<\rho_{2N}\:,\ \vert z_0\vert>0$ can cancel each other.

\subsubsection{3D Impurity States, Low QR}

It is clear that the binding energy $E_b=-\Delta W$ (\ref{E:bdelW}) increases
with a decrease of the height $d<a_0$ of the QR and with increasing
strength of the electric field strength. For small displacements $2z_0/d\ll1$
from the mid-plane $z_0=0$ we obtain from (\ref{E:bdelW})
\begin{equation}\label{E:d1}
E_b(z_0)=E_{b1}(0) + 2 Ry \left(\frac{a_0}{d}\right)
\!\! \left(\frac{2z_0}{d}\right)^{\!2\!}
\left\{
- 1 + \pi^2 \left[\:\! C \!-\! \Delta(\rho_0) \!+\!
\ln\!\left(\frac{\pi\rho_{2N}}{d}\right) \right]
\right\}
\end{equation}
where
$$
E_{b1}(0)=-\:\!8\:\! Ry\left(\frac{a_0}{d}\right)
\left[ C-\Delta(\rho_0) +
\ln\left(\frac{\pi^{1/2}\rho_{2N}}{d}\right) \right]
$$
is the binding energy of the impurity positioned at the mid-plane $z_0=0$ and
any radial distances $\rho_0$. The shift of the impurity from the point $z_0$
leads to a decrease with respect to the binding energy. With increasing electric field strength
we observe an increase of the shift of the binding energy caused by the displacement
$z_0$.

The dependence of the binding energy on the radial position $\rho_0$ can be
derived from eq.(\ref{E:bdelW})
\begin{equation}\label{E:d2}
E_b(\Delta)=E_{b2}(0) + 8\:\! Ry
\left(\frac{a_0}{d}\right) \Delta\left(\rho_0\right)
\cos^2\left(\frac{\pi z_0}{d}\right)
\end{equation}
where
$$
E_{b2}(0)=2Ry\left(\frac{a_0}{d}\right)
\left\{ -\:\! 4 \left[\:\! C + \ln\left(\frac{\pi \rho_{2N}}{d}\right) \right]
\cos^2\left(\frac{\pi z_0}{d}\right) +
\ln\pi^2\left(1-\frac{4 z_0^2}{d^2}\right)\right\}
$$
is the binding energy of the impurity positioned at the point $\rho_0=\rho_{2N}$
for which $\Delta(\rho_{2N})=0$ and at any plane $z_0$. The binding energy
$E_b$ (\ref{E:d2}) decreases if the impurity center moves from the certain
position $\rho_{0m}$
towards the radial boundaries of the QR. The shift of the binding energy
associated with the parameter $\Delta(\rho_0)$ reaches a maximum for the
impurity center positioned at the plane $ z_0=0$ and decreases with increasing
displacement from this plane. For a low QR the corrections to the binding
energy induced by the radial $(\rho_0)$ and vertical $(z_0)$ displacements
can be chosen such that they cancel each other, i.e. in this case there is no
resulting change of the energy for specific shifts from the circle
$\rho_0\simeq\rho_{2N}$, $z_0=0$ to the region $a\leq\rho_0<\rho_{2N}$,
$|z_0|>0$. Thus the dependencies of the impurity binding energy $E_b$
on the strength of the electric field, the height of the QR $d$ and the
position of the impurity within the QR $\rho_0, z_0$ are qualitatively the same
both for high and low QRs.

\section{Numerical Approach}

\subsection{Computational Method}

Our numerical approach to solve eq.(\ref{E:basicEq}) is a
finite difference method, described in detail in Refs. \onlinecite{Ivanov86,Ivanov88,IvaSchm2001b}
for two-dimensional systems and in Refs. \onlinecite{Ivanov97,IvaSchm2002} for three-dimensional
systems. We have solved eq.(\ref{E:basicEq}) in cylindrical coordinates
$(\rho,\phi,z)$ in a region $\Omega$
\begin{eqnarray}
a\leq\rho\leq b\nonumber\\
0\leq\phi\leq\pi\nonumber\\
-d/2\leq z\leq d/2
\label{eq:numregion}
\end{eqnarray}
respecting the boundary conditions (\ref{E:Cond}), i.e. with $\Psi=0$ on
the boundaries in coordinates $\rho$ and $z$ and with
condition $\left.\frac{\partial\Psi}{\partial\phi}\right|_{\phi=0,\pi} =0$
for $\phi$. Our computational procedure consists of the following main steps.
The nodes of the spatial mesh are chosen in the domain $\Omega$ and the values of the wave function at
the nodes represent solutions of the initial differential equation (\ref{E:basicEq}).
Since the domain $\Omega$ is bounded with respect to all three coordinates we can use
uniform meshes. The nodes of these meshes have coordinates
$\rho_i=a+(b-a)(i-1/2)/N_\rho$,
$\phi_j=\pi(j-1/2)/N_\phi$, and
$z_k=-d/2+d(k-1/2)/N_z$, $i=1,\ldots,N_\rho$,
$j=1,\ldots,N_\phi$, $k=1,\ldots,N_z$.
After replacing the derivatives by their
finite-difference approximations eq.(1) takes a form of
a system of linear equations for $\Psi$ values at the
nodes, and approximate values of energy can be found
as eigenvalues of the corresponding Hermitian matrix.
The final values for the energy are provided by using
the Richardson extrapolation technique for the corresponding
results emerging from a series of geometrically similar
meshes with different number of nodes, i.e.
eigenvalues obtained for meshes with
$N_\rho=KN_{\rho 0}$, $N_\phi=KN_{\phi 0}$, $N_z=KN_{z 0}$,
where $K=1,2,\ldots$.
Using this approach we achieve a major increase of the
numerical precision and, in particular, we obtain together with
each numerical value a reliable estimate of its precision.
Typical numbers of mesh nodes used in the present calculations
were of order $40^3$, i.e. 40 nodes in each direction for
the thickest (corresponding to maximal values of $K$)
meshes.  An important factor affecting the choice of values $KN_{\rho 0}$,
$KN_{\phi 0}$, and $KN_{z 0}$ is the position of the Coulomb
center, which does not coincide with the origin of the
coordinate system.
A geometrical similarity of meshes with different $K$
can be achieved only when this center has coordinates
$(a+(b-a)i/N_{\rho 0},\ \pi j/N_{\phi 0},\ -d/2+dk/N_{z0})$.
This circumstance affects the choice of coordinates of
the Coulomb center and sometimes required calculations
on meshes containing more nodes than absolutely needed for
obtaining a satisfactory numerical precision.
Along with solutions of eq.(\ref{E:basicEq}) in its general 3D form we have
solved the corresponding equation in two dimensions employing the coordinates $(\rho,\phi)$
($z=z_0$). This allows us obtaining the binding energy of the electron
in the limit $d\rightarrow 0$. The numerical solution has no additional specific features
compared to the 3D one.

\subsection{Numerical Results and Discussion}

We present here detailed results on the binding energy of
the electron for two fixed geometries of the quantum ring,
referred to in the following as (A) and (B)
corresponding to realistic experimental parameters (see below). Some
additional results on the dependencies of this energy on
the geometry of the quantum ring are also presented.
Parameters of the quantum ring (A) are:
GaAs is the ring material, GaAlAs is the barrier material,
$a=5$~nm, $b=20$~nm, and $d=15$~nm.
For the quantum ring (B): $a=10$~nm, $b=40$~nm, and $d=3$~nm with InAs as the ring
material and GaAs as the barrier material.
In order to simplify the comparison of the results for
rings made of different materials we transform the values of
parameters into effective atomic units (e.a.u.).
Using parameters $\epsilon=12.56$ and $\mu=0.067m_0$ for GaAs and
$\epsilon=14.5$ and $\mu=0.023m_0$ for InAs we obtain
$a=0.5$, $b=2$, and $d=1.5$ for ring (A) and
$a=0.3$, $b=1.2$, and $d=0.09$ for (B).
The case (A) means a ring with its height being comparable with the other dimensions,
whereas case (B) is a low ring with $d\ll a,b$.

\begin{figure}
\includegraphics[width=8.5cm,clip]{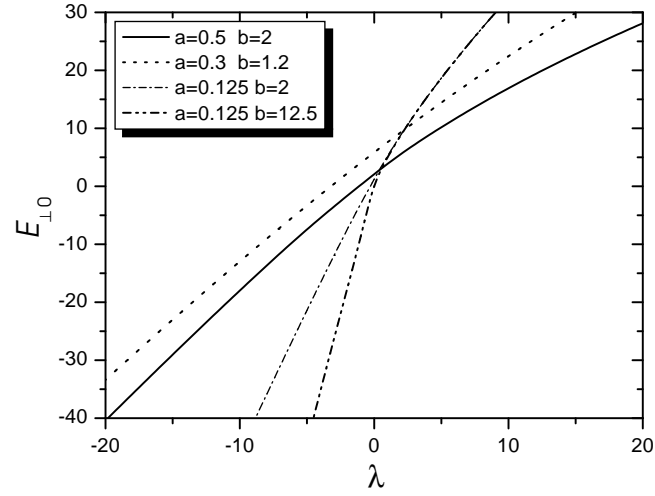}
\caption{Energy of the lateral motion a free electron in a quantum ring as a function
of the charge $\lambda$ of the central wire. Effective atomic units are used.
\label{figur0}}
\end{figure}

In the following we will particularly consider the binding energy $E_{\rm b}$ of the electron
as a function of various parameters.
The binding energy is the difference between the total energy of the electron
obtained in our numerical calculations and the energy $E_0$ of the electron
in the same quantum ring without the impurity center .
The latter energy consists of two terms
\begin{eqnarray}
E_0=E_{\| 0}+E_{\perp 0}
\label{eq:zeroen}
\end{eqnarray}
where $E_{\| 0}$ and $E_{\perp 0}$ are the energies of the motion
in the $z$ direction and of the lateral motion in the plane
$(\phi,\rho)$ respectively.
An exact analytical expression for the first of them is given
in eq.(8) and for the ground state looks for effective atomic units as
$E_{\| 0}=\frac{\pi^2}{2d^2}$.
The second term does not depend on $d$ and can be easily calculated
numerically for each set of parameters $(a,b,\lambda)$ by solving
a two-dimensional version of eq.(1) for very large values of $\rho_0$.
In Figure~\ref{figur0} we show $E_{\perp 0}(\lambda)$
for several geometries.
They include $E_{\perp 0}(\lambda)$ for QR's (A) and (B) as well as two
dependences for a small inner radius $a=0.125$~e.a.u. and different $b$.
For large positive $\lambda$ these functions are near to linear ones
owing to the concentration of the electron density in a small vicinity
of the inner boundary of the quantum ring.
In result the curves for $a=0.125$~e.a.u. and different $b$ coincide
for this range of $\lambda$.
The slope of the curves for positive $\lambda$ is determined by $a$
and increases with a decrease of this value.
For large negative $\lambda$ the functions $E_{\perp 0}(\lambda)$ also
have a linear form with slopes depending on both $a$ and $b$.

\begin{figure}
\includegraphics[width=8.5cm,clip]{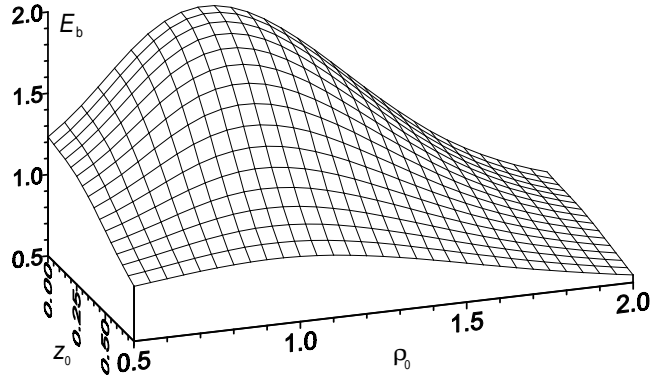}
\caption{Binding energy $E_{\rm b}$ as a function of the position of the impurity center for
quantum ring (A) $\lambda=4.3$
of the central wire. Effective atomic units are used.
\label{figur1}}
\end{figure}

In Figure~\ref{figur1} we present {$E_{\rm b}$ as a function of the position
$(z_0,\rho_0)$ of the impurity center for quantum ring (A).
The binding energy shows a significant dependence on $\rho_0$ and $\lambda$ for
$z_0\ll d/2$ whereas for $z_0$ being close to $\pm \frac{d}{2}$ the dependency
on $\rho_0$ is much less pronounced.
The dependence of $E_{\rm b}$ on $\rho_0$ and $\lambda$
for $z_0=0$ is presented in Figure~\ref{figur2}.
It should be noted that the effect of the radial electric field
on the binding energy is much less compared to its effect on the
total energy.
For example, the maximal value of the difference
$E_{\rm b}(\lambda=10)-E_{\rm b}(\lambda=0)$ (for $\rho_0=0.875$)
is 0.495~e.a.u., whereas the corresponding difference for
$E_0$ (or $E_{\perp 0}$) is 14.81~e.a.u. However, the relative change of $E_{\rm b}$ due to
the external field is of the order of $100 \%$.

\begin{figure}
\includegraphics[width=8.5cm,clip]{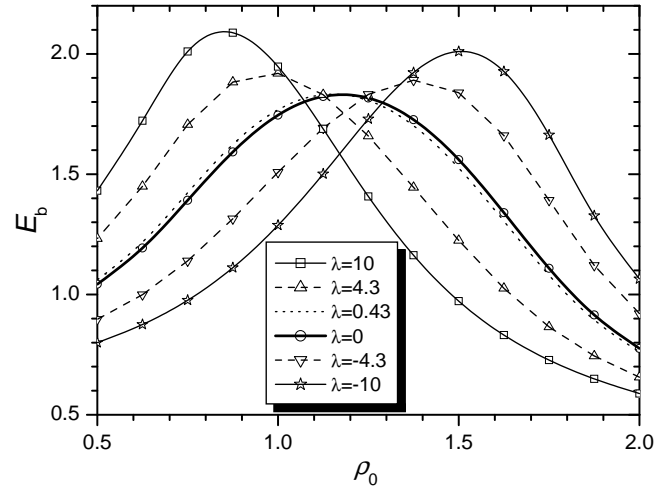}
\caption{Binding energy $E_{\rm b}$ as a function of the radial position
of the impurity center $\rho_0$ for
quantum ring (A) for $z_0=0$ for several different values
of the linear charge density $\lambda$
of the central wire. Effective atomic units are used.
\label{figur2}}
\end{figure}

For $\lambda=0$ the binding energy presented in
Figure~\ref{figur2} achieves its maximum (at $\rho_{0m}$) close to the middle point
of the radial cross-section of the quantum ring.
This is due to the fact that the energy of the ground state of a hydrogen-like system increases
when the system approach a impenetrable potential wall.
In the case of a flat infinite wall the ground state
of a hydrogen atom with the nucleus lying on the boundary
of the wall is similar to the state $2p_0$ of the
free hydrogen atom and has energy $-0.125$~a.u.
instead of $-0.5$~a.u. for the ground state of
the free atom. Since $E_0$ does not depend on the position
of the impurity center approaching the boundaries
of the quantum ring equally affects the value of the total energy and
the value of the binding energy $E_{\rm b}$.
On the other hand, positive values of $E_0$ for spatially
confined systems ($\lambda=0$) increase the binding energy
compared to a free impurity center.
As a result the impurity electron in the quantum ring is more tightly
bound for all the parameter values presented in Figure~\ref{figur2}
compared to the case of an impurity center in a bulk.
However, the total energies of the electron are higher compared to the case of a bulk.

The asymmetry of the curve for $\lambda=0$ and in particular a higher binding
energy at $\rho_0=a$ compared to $\rho_0=b$ are due
to the curvature of the boundaries of the quantum ring.
In the case of the inner $\rho_0=a$ boundary its curvature provides
more space for the motion of the electron compared to a corresponding flat wall.
In result the motion of the electron is less confined
and its energy is lower than for the case of a flat wall
and is closer to the values for the case where the center is far from
the boundaries. The opposite curvature of the outer $\rho_0=b$ boundary
of the quantum ring leads to the opposite effect
consisting of a decrease of the binding energy
for the impurity center near to this boundary.

For both positive and negative $\lambda$ the maxima
of the energy curves $E_{\rm b}(\rho_0)$ exceed that
corresponding to $\lambda=0$.
This effect is fully analogous to the quadratic Stark effect,
which leads to the decrease of the ground state energy level and
to the increase of the binding energy of the hydrogen-like impurity
electron in the presence of external electric field.
The shifts of the positions of the maximums to
the left hand side for $\lambda>0$ and to the
right hand side for $\lambda<0$ are explained by the
increase of the binding energy for smaller distances $\rho_0$
in case of an attractive positively charged wire or
for larger distances in case of a negatively charged wire.

Figure 3 (curves corresponding to $\lambda = \pm 4.3$ ) demonstrates that
in the impurity QR the inversion effect of the electric field occurs.
The binding energy $E_b (\lambda)$ changes as the direction of the
electric field $\lambda $ changes ($+\lambda \to -\lambda$). An analogous
effect relating to a quantum well structure was studied in Ref. \onlinecite {Mon}.
In contrast to the quantum well in which the inversion shift of the binding energy
$\Delta E_b=E_b (+\lambda)-E_b(-\lambda)$ vanishes for the impurity centre
positioned at the mid-plane ($z_0=0$) , the inversion shift
in the QR is absent for a certain ($\rho_0 = 1.15$) cylindrical surface.

\begin{figure}
\includegraphics[width=8.5cm,clip]{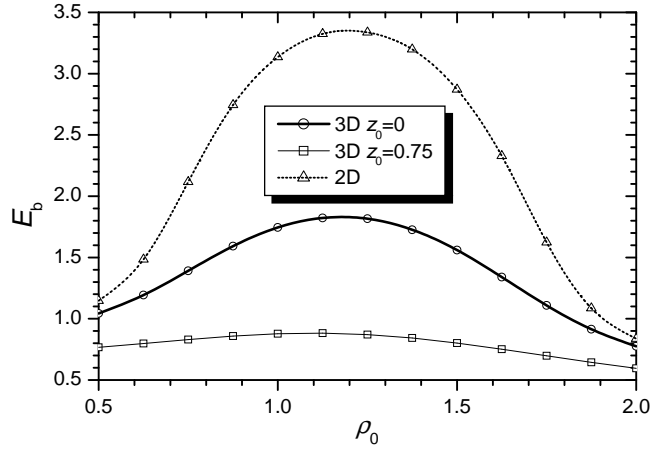}
\caption{Binding energy $E_{\rm b}$ as a function of the radial position
of the impurity center $\rho_0$ for
quantum ring (A) for $\lambda=0$ and for different displacements $z_0=0$,
$z_0=d/2$
and the corresponding 2D quantum ring ($d=0$).
Effective atomic units are used.
\label{figur3}}
\end{figure}

\begin{figure}
\includegraphics[width=8.5cm,clip]{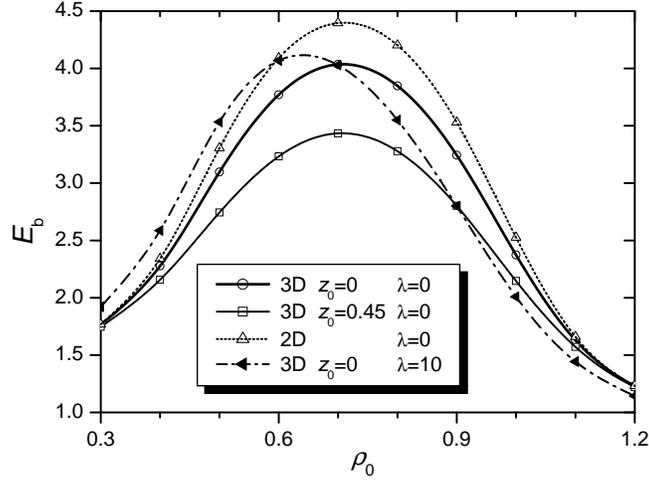}
\caption{Quantum ring (B): same as Figure~\ref{figur3}
and a 3D curve for $\lambda=10$, $z_0=0$. Effective atomic units are used.
\label{figur4}}
\end{figure}

\begin{figure}
\includegraphics[width=8.5cm,clip]{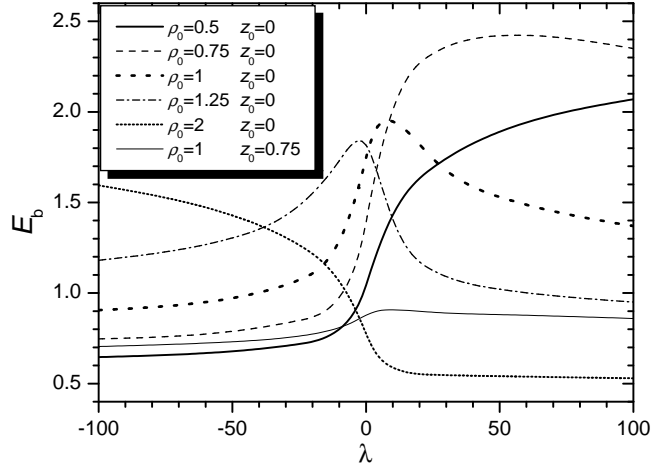}
\caption{Binding energy $E_{\rm b}$ as a function of the central wire charge
$\lambda$ for the different impurity positions $\rho_0$, $z_0$ in the quantum
ring (A). Effective atomic units are used.
\label{figur5}}
\end{figure}

In Figure~\ref{figur3} we compare the binding energies $E_{\rm b}(\rho_0)$
for the quantum ring (A) for $z_0=0$, $z_0=d/2$, and for
a 2D quantum ring with the same values of $a$ and $b$.
We observe a much weaker dependence of the binding energy for $z_0=d/2$
on the radial position of the impurity center compared to the case $z_0=0$.
On the other hand, the curve for $z_0=0$ is much lower in energy
and demonstrates a weaker dependence on $\rho_0$ compared to the 2D curve.
The large difference between 2D and 3D binding energies is due to the
large value of $d$ for the quantum ring (A) and strong confinement provided
by the 2D QR.
The predominant part of the curve $E_{\rm b}$ for the two-dimensional impurity is above
$E_{\rm b}=2$. The latter is the binding energy for a two-dimensional impurity
without external confinement for the motion in $\rho$-direction i.e. a bulk impurity.
The above-mentioned difference reflects the increase of the binding energy
due to the confinement in the radial $\rho$-direction.

The opposite case of a low quantum ring is shown in
Figure~\ref{figur4} where the corresponding energy curves analogous to those
of Figure~\ref{figur3} are presented for the quantum ring (B).
For $\lambda=0$ the behaviour is qualitatively very
similar to the one observed in Figure~\ref{figur3} with the exception that both 3D energy
curves are much closer to the 2D curve than for the quantum ring (A).
Figure~\ref{figur4} contains also an energy curve
for a non-zero radial electric field, corresponding
to $\lambda=10$.
One can see that the radial electric field has a minor
effect for the low QR (B) compared to the high QR (A).
This is due to the quantitatively reduced effect of an external electric field
on 2D hydrogen like systems compared to 3D ones
(see Ref.\onlinecite{D0Fpap,2D0Fpap} and references therein).

The presence of maxima for the energy functions
$E_{\rm b}(\rho_0)$ and the shifts of these maxima
with changing radial electric field strength determine the form of these
functions $E_{\rm b}(\lambda)$ presented in Figure~\ref{figur5}.
For the impurity being centered in $z$-direction
($z_0=0$) we present five curves for different positions $\rho_0$
in the radial direction. Let us discuss some major properties
of these curves. It is straightforward to understand the behaviour of the two curves for $\rho_0=1$ and $\rho_0=1.25$
corresponding to the impurity center positioned near to the middle of the QR in the
radial direction.  These curves have maxima in the vicinity of $\lambda=0$.
This means that the presence of a strong radial electric
field (independent of its direction) decreases the binding energy
of the electron to the impurity center for these values of $\rho_0$.
This decrease of the binding energy originates from a decrease
of the electronic density near the impurity center, because
the radial electric field either attracts the electronic density to
the inner boundary of the ring or repels it towards the outer
boundary. The curves $E_b(\lambda)$ for the two opposite cases when the impurity center is situated
directly on the inner $\rho_0=0.5$ or on the outer $\rho_0=2$
boundary of the QR possess no maxima. They show an increase of the binding energy when the radial
electric field attracts the electronic density to the impurity
center and a decrease of $E_{\rm b}$ when it repels the electronic
density to the opposite boundary.
If the boundaries would be flat and infinite in the $z$ direction
the limiting values of the binding energy both for $\rho_0=0.5$ and
$\lambda\rightarrow+\infty$ and for $\rho_0=2$ and
$\lambda\rightarrow-\infty$ would correspond to the
ground state of a two-dimensional hydrogen atom $E_{\rm b}=2$
(the mutual action of the electric field and the impenetrable
potential wall would provide a confinement of the electron in the plane).
In the QR the outer boundary provides less space for the motion of
the electron in the corresponding limit in comparison with
the inner boundary.
In result we may expect that the limiting value of $E_{\rm b}$
for  $\rho_0=0.5,\ \lambda\rightarrow+\infty$ should be larger than
$E_{\rm b}$ for  $\rho_0=2,\ \lambda\rightarrow-\infty$.
This is in agreement with the behavior of the corresponding curves in Figure~\ref{figur5}.
For the two opposite limits  $\rho_0=0.5,\ \lambda\rightarrow-\infty$ and
$\rho_0=2,\ \lambda\rightarrow+\infty$ $E_{\rm b}(\lambda)$
becomes asymptotically flat.
This is due to the fact that when practically all the electronic density
is concentrated at the opposite boundary its small redistribution
in strong fields does not affect
(due to different but large $\lambda$ values)
the interaction with the impurity center.
The energy of this interaction depends on the average distance
between the electron and the impurity center.
Due to the confinement of the electron on the inner or outer
surface of the quantum ring the latter problem acquires some similarity
with the electrostatic problem of a charge near a conducting surface.
The potential of the interaction of a charge with its image
in a convex surface is smaller than the interaction of the charge
with a mirror charge in a flat or a concave surface
(compare with ref.~\onlinecite{Jackson}).
These reasons explain the result
$E_{\rm b}(\rho_0=0.5,\ \lambda\rightarrow-\infty)>
E_{\rm b}(\rho_0=2,\ \lambda\rightarrow+\infty)$.

As an example of $E_{\rm b}(\lambda)$ for an impurity center
located at a small distance from one of the boundaries
we show in Figure~\ref{figur5} a curve for $\rho_0=0.75$
and $z_0=0$.
In agreement with the picture presented above it has a maximum
in a region of positive $\lambda$.
The binding energy in the vicinity of this maximum is higher than
that for maxima of the corresponding curves for $\rho_0=1$ and $\rho_0=1.25$ because
the radial electric field can concentrate a larger electronic
density at smaller $\rho$.
Finally we present also $E_{\rm b}(\lambda)$ for $\rho_0=1$ and
$z_0=0.75$, i.e. for the impurity being located on the top boundary
of the QR.
The behavior of this curve is similar to curves for $z_0=0$,
but both the absolute values of $E_{\rm b}$ and their alterations
are smaller. Note that in the QR the radial electric field may provide the
impurity electron to be more stable while in the bulk material the electric field
leads to the ionization of the impurity centre.

The obtained results allow to estimate the values to be expected in an experiment.
It follows from Figure 6 that the positive shift of the binding energy
$\Delta E_b (\lambda)$of the impurity positioned at $z_0=0$ and $\rho_0=7.5$nm
in the QR(A) caused by a positive electric field $\lambda$ of a charged wire
having the linear electron density $n_e=5 \cdot 10^6$~cm$^{-1}$ amounts to
$\Delta E_b \simeq 4.5$~meV. The relative change is therefore
$\frac{\Delta E_b (\lambda)}{E_b (0)}=0.28 $. In this electric field a
decrease of the binding energy $\Delta E_b\simeq - 11.6$~meV
(about 52 percents) occurs if we shift the impurity centre from the
symmetric plane $\rho_0 = 10$~nm, $z_0=0$ to the boundary plane $\rho_0=10$~nm,
$z_0=7.5$~nm. When the impurity centre moves in the symmetric plane ($z_0=0$)
of the QR(A) (see Figure 3) subject to the positive electric field $\lambda=4.3$
determined by the linear electron density $n_e=4.3 \cdot 10^6$~cm$^{-1}$~
from the internal boundary~ $\rho_0=5$nm to the position $\rho_0=10$~nm the
binding energy
$E_b$ increases by an amount $\Delta E_b\simeq 8.1$~meV. This is about 58
percents of the binding energy $E_b(\rho_0)$ at $\rho_0=a$. In this field
the inversion shift of the binding energy
$\Delta E_b(\lambda)=E_b(-\lambda)-E_b(+\lambda)$ for the case $z_0=0$,
$\rho_0=15$~nm is $\Delta E_b\simeq 7.5$~meV. The inversion shift vanishes at
$\rho_0=11.5$~nm. The estimates of the expected values for the InAs/GaAs QR
can be made accordingly using the parameters of the InAs material, providing
values of the same order of magnitude as those for the GaAs QR.
Thus the obtained effects induced by the radial electric
field in the impurity QR are detectable in an experiment.

Concerning the currently available experimental
data to our knowledge most of them are related to the persistent
current occuring in the QR threaded by the magnetic field (see Ref. \onlinecite{LW02}
and references therein). The effect of an electric field is studied
theoretically \cite{Ll09,Ba07,LW02} for the case of a uniform electric field
directed parallel to the plane of the QR. One of the reason (concerning both
theory and experiment) to choose this configuration is that
the electric field has been treated as a tool acting
on the electron states causing in particular the breaking of the axially symmetric
potential of the QR and mixing the states with different angular quantum numbers. 
No additional serious experimental refinements
specifically associated with the QRs are required.
However the radially directed electric field considered in our paper is capable on the
one hand to conserve the axial symmetry of the QR potential and on the other hand
to  modify strongly the impurity states in the QR.
Let us briefly address the question of the experimental realization of the additional electric field.
A Si wire covered by the
exact-position monitored charge being embedded in the inner region of the QR
offers to be a source of the radial electric field. Particularly these wires
of about 10 nm diameter are employed in the silicon-based charge-coupled
devices \cite{Fuji}. An alternative approach would be as follows. The inner region of the QR is doped
by the shallow impurity centres. Being activated these centres become
charged and this region can be treated as the source of the radial
electric field for the QR. We believe that our results could stimulate
experiments that contribute to the physics of QRs and their
optical-electronic applications.

\begin{figure}
\includegraphics[width=8.5cm,clip]{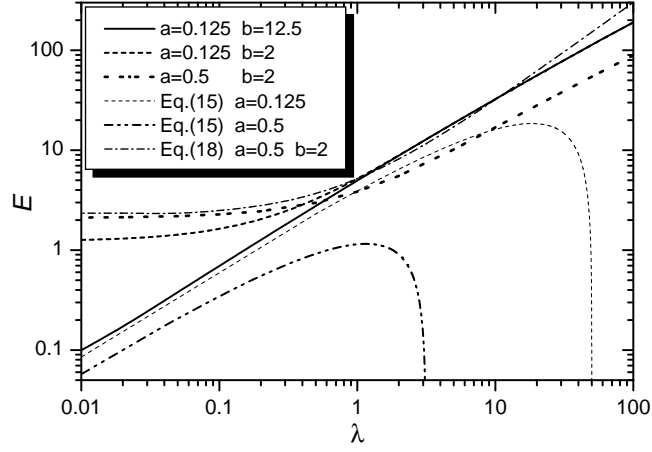}
\caption{Energy of a free electron in three different two-dimensional
quantum rings as a function of the positive charge of the central wire $\lambda$.
Numerical results and analytical estimations are shown. Effective atomic units are used.
\label{figur6}}
\end{figure}

\begin{figure}
\includegraphics[width=8.5cm,clip]{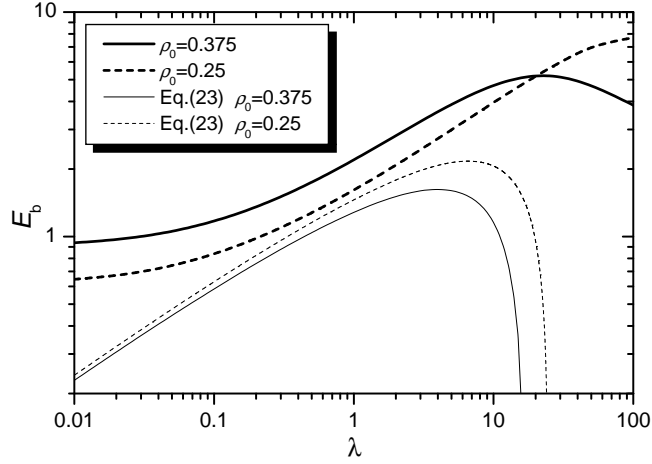}
\caption{Binding energy of an electron to an impurity center in a two-dimensional
quantum ring as a function of the charge of the central wire.
Numerical results and analytical estimates are shown.
Effective atomic units are used.
\label{figur7}}
\end{figure}

\section{Comparison of Numerical and Analytical Results}

We present in this section an exemplary comparison
of our numerical and analytical results.
Figure~\ref{figur6} shows numerically calculated energies for a free electron
in three different two dimensional quantum rings.
These energies are given as functions of the logarithm of
the positive charge density on the central wire.
For a QR with small $a=0.125$ and very large $b=12.5$
the numerical curve for $E(\lambda)$ presented on a double
logarithmic scale is not very different from a straight line.
The formula (15) provides a good approximation for this curve
for not too large values of $\lambda$.
The discrepancy at large $\lambda$ is due to small values of
$\rho_{2N}$ ($\rho_{2N}<a$) given by eq.(15).
A major improvement of the analytical estimation could be
achieved by replacing the formula for $\rho_{2N}$ by
\begin{eqnarray}
\rho_{2N}=\left(\frac{4\pi}s\right)^{1/2}(N+1/2)+a
\label{eq:newrho2n}
\end{eqnarray}

The numerically calculated curve for $a=0.125$ and $b=2$ coincides
with that for $a=0.125$ and $b=12.5$ for large values of $\lambda$.
As $\lambda\rightarrow 0$ the value of $E$ converges to a finite
limit, determined by the energy of an electron in a finite 2D
quantum ring.
This energy increases when the ring becomes narrower as can be
seen when comparing the curves for $a=0.125$ and $b=2$ and
for $a=0.5$ and $b=2$.
For large $\lambda$ the latter curve shows also approximately a linear behaviour
but is lower in energy than the curves for $a=0.125$.
This shift is properly described by eq.(15), but the relatively small value
$b=2$ does not allow to obtain estimations for the energy of
the free electron for $a=0.5$ and $b=2$ by eq.(15).
For this set of parameters we present in Figure~\ref{figur6}
also the energy given by eq.(18).
One can see that even for such a broad QR this formula
gives quite a reasonable approximation to the energy.
For $(b-a)\ll a,b$ this formula is in very good agreement
with the numerical data.

In Figure~\ref{figur7} analytical estimates given by eq.(23)
for the binding energy of an electron in two-dimensional quantum rings
in the presence of an impurity are compared with the
corresponding numerical results. Eq.(23) includes $\rho_{2N}$ values given by eq.(15) and this
circumstance restricts the applicability of it for strong fields.
Condition $\rho_{2N}\ll a_0$, or $\rho_{2N}\ll 1$ in effective
atomic units, restrict the applicability of (23) to weak fields.
Within these conditions the agreement of the estimates
and numerical results is good. This is visible for the curves $\rho_0=0.25$,
which fulfill the conditions for the validity of eq.(23) in an improved manner
compared to the case $\rho_0=0.375$.

\section{Summary and Conclusion}\label{S:Concl}

We have studied analytically and numerically the problem of an
impurity electron in a QR in the presence of a radially directed external
electric field. The two-fold character of our investigation illuminates the physical
behaviour and properties of our impurity/quantum ring system in a complementary way.
The basis of the analytical approach is an adiabatic quasiclassical
approximation, while a finite-difference method in two and three dimensions
was used to perform the numerical calculations.
For our analytical studies the external electric field is taken
to be much stronger than the electric field due to the interaction with the impurity.
The dependencies of the binding energy of the impurity electron on the strength of the
external electric field, the parameters of the QR and the position of the
impurity center within the QR are derived explicitly.

We have shown that if the height of the QR increases and/or the impurity
center displaces  from the mid-circle ($z_0\neq 0$) for any radius of the QR towards
the boundary planes $z_0=\pm d/2$ of the QR, the binding
energy decreases.  The binding energy reaches a maximum for the impurity
positioned at the mid-plane $z_0=0$ of the QR. For a fixed $z_0$ and
without the radial electric field the binding energy has its maximum
close to the middle point of the radial cross-section of the quantum ring.
The radial electric field shifts the position of the maximum
towards the center of the ring in case of a positive charge of the
central wire $\lambda$ and in the opposite direction for a negative charge.
This results in a relatively complicated dependence $E_{\rm b}(\lambda)$,
which is very different for different distances of the impurity from the center of the QR.
The maximum value of the binding energy increases with increasing electric field strength.
The amplitudes of the mentioned dependencies decrease while shifting
the impurity towards to the boundary planes of the QR. The inversion effect i.e.
the change of the binding energy when the direction of the electric
field is changed to the opposite one is realized in the impurity QR.
Estimates of the binding energies for realistic strengths of the external electric
field and the parameters of GaAs and InAs quantum rings are provided.

We have demonstrated that a strong radial electric field and ring confinement provide
considerable polarization phenomena of the impurity states.
Strong dependencies of the binding energy of the impurity electron 
should lead to significant changes of transport processes
and optical properties of the QRs. The great sensivity of the impurity QRs
to the radial electric field is useful for its applications in field-effect
transistor structures, electrooptical modulators and switching devices.

\end{document}